\documentstyle[11pt]{article}
\title{On the Uniqueness of the Chentsov Metric in Quantum
Information Geometry}
\author{M. R. Grasselli\thanks{Supported by a grant from CAPES-Brazil.}
\qquad\qquad R. F. Streater \\Dept. of Mathematics,\\King's College,
\\Strand,\\London, WC2R 2LS}
\date{11/06/2000}

\newtheorem{theorem}[equation]{Theorem}

\begin{document}
\setlength{\evensidemargin}{.1in}
\setlength{\oddsidemargin}{.1in}
\maketitle
\begin{abstract}
We show that, in finite dimensions, the only monotone metrics on
the space of invertible density matrices for which the $(+1)$ and $(-1)$
affine connections are mutually dual are constant multiples of the
Bogoliubov-Kubo-Mori metric.
\end{abstract}

\section{Introduction}

The first task of quantum information geometry is to extend to
non-commutative probability spaces the uniqueness and rigidity of
the geometrical structures used in the classical version of the
theory. This includes, for instance, the uniqueness result of Chentsov concerning
the Fisher metric \cite{Chentsov} and the two equivalent
definitions of $\alpha$-connections given by Amari \cite{Amari1}:
either using an $\alpha$-embedding of the form $p \mapsto
\frac{2}{1-\alpha}p^{\frac{1-\alpha}{2}}$ or as the convex mixture
of the exponential and mixture connections.

As for the first problem, Chentsov himself made the first attempt
to find the possible Riemannian metrics on a quantum information
manifold with the property of having its line element reduced
under stochastic maps \cite{MorozovaChentsov} (cf \cite{Petz3}). This monotonicity
property, which is the quantum analogue of being reduced under Markov morphisms,
was later investigated by Petz \cite{Petz3}. Unlike the
classical case, he found that there are infinitely many Riemannian metrics
satisfying it. Concerning the second problem, several definitions
of $\alpha$-connections have been proposed
\cite{Nagaoka,Hasegawa1,GibiliscoIsola}, both for finite
and infinite dimensional quantum systems. Since some of these
definitions involve finding  dual connections with respect to
some chosen Fisher metric, it is clear that the multitude
of possible candidates for the metric encourages the appearance of
non-equivalent definitions for the $\alpha$-connections.

We take the position that quantum information manifolds are
equipped with two natural flat connections: the mixture
connection, obtained from the linear structure of trace class
operators themselves, and the exponential connection, obtained
when combinations of states are performed by adding their
logarithms \cite{GrasselliStreater,Streater}. In \S2 we present our
preferred definition of the
$\pm 1$-connections in finite dimensional quantum information manifolds.

Following Amari \cite{Amari1,Amari2}, we consider duality as a fundamental structure
to be explored. Thus, given these two connections, we should ask what are the Riemannian
metrics that make them dual. We give our answer to this question in \S3.
In \S4, we combine this result with Petz's characterisation
of monotone metrics in finite dimensional quantum systems to
find that the {\em BKM} metric \cite{Bogoliubov,Kubo,Mori} is, up to a
factor, the unique monotone Riemannian metric with respect to
which the exponential and mixture connections are dual.

\section{The Exponential and Mixture Connections}

Let ${\cal H}^N$ be a finite dimensional complex Hilbert space, $\cal A$
the subspace of self-adjoint operators and $\cal M$ the set of all
invertible density operators on ${\cal H}^N$. Then $\cal A$ is an
$N^2$-dimensional real vector space and $\cal M$  is
an $n$-dimensional submanifold with $n=N^2-1$. Defining
the $1$-embedding of $\cal M$ into $\cal A$ as
\begin{eqnarray*}
\ell_1 & : & {\cal M} \rightarrow{\cal A}\\
& & \rho \mapsto \log \rho,
\end{eqnarray*}
we can use the linear space structure of $\cal A$ to obtain a representation
of the tangent bundle of $\cal M$ in terms of operators in $\cal A$ even
though $\cal M$ is not a vector space itself. At each point $\rho \in {\cal M}$,
consider the subspace ${\cal A}_{\rho}=\{A \in {\cal A} : \mbox{Tr} (\rho A) =
0 \}$ of $\cal A$, called the space of `scores'; we define the isomorphism
\begin{eqnarray*}
(\ell_1)_{*(\rho)} & : & T_{\rho}{\cal M} \rightarrow {\cal A}_{\rho} \\
& & v \mapsto (\ell_1 \circ \gamma)^\prime (0),
\end{eqnarray*}
where $\gamma :(-\varepsilon,\varepsilon) \rightarrow {\cal M}$ is a
curve in the equivalence class of the tangent vector $v$. We call this
isomorphism the
$1$-representation of the tangent space $T_{\rho}{\cal M}$. If
$(\theta^1,\ldots ,\theta^n)$ is a coordinate system for $\cal M$,
put $\partial_i:=\partial/\partial\theta^i,\;i=1,\ldots,n$, evaluated
at $\rho$. Then the $1$-representation of the basis $\left\{\partial_1,
\ldots ,{\partial_n}\right\}$ of $T_{\rho}{\cal M}$ is
$\left\{ \frac{\partial\log\rho}{\partial\theta^1},
\ldots ,\frac{\partial\log\rho}{\partial\theta^n}\right\}$. The
$1$-representation of a vector field $X$ on $\cal M$ is therefore
the $\cal A$-valued function $(X)^{(1)}$ given by
$(X)^{(1)}(\rho)=(\ell_1)_{*(\rho)}X_{\rho}$.

The exponential or $1$-connection is the connection obtained
from the $1$-embedding through the following parallel transport
\cite{Streater}
\begin{eqnarray*}
\tau^{(1)}_{\rho_0,\rho_1} & : & T_{\rho_0}{\cal M} \rightarrow T_{\rho_1}{\cal M}
\\                   &   & v \mapsto
(\ell_1)_{*(\rho_1)}^{-1}\left((\ell_1)_{*(\rho_0)}v -
\mbox{Tr}[\rho_1(\ell_1)_{*(\rho_0)}v]\right).
\end{eqnarray*}
Giving the parallel transport in a neighbourhood of $\rho$ is equivalent
to specifying the
covariant derivative. It is readily verified that the $1$-representation of
the $1$-covariant derivative, applied to the vector field $\partial_j:=
\partial/\partial\theta^j$, is
\begin{equation}
\label{exp-cov}
\left(\nabla^{(1)}_{\partial_i}\frac{\partial}
{\partial\theta^j}\right)^{(1)}=
\frac{\partial^2\log\rho}{\partial\theta^i\partial\theta^j}-
\mbox{Tr}\left(\rho\frac{\partial^2\log\rho}{\partial\theta^i\partial
\theta^j}\right)
\end{equation}

At a more abstract level, the construction above corresponds to making
${\cal M}$ into an affine space and endowing it with the {\em natural} flat
connection \cite{Murray}. Rather than exploring this
concept further, we benefit from dealing with finite dimensional
spaces and prove that the $1$-connection is flat by exhibiting an affine
coordinate system for it. Let $\left\{{\bf{1}}, X_1, \ldots, X_n\right\}$ be
a basis for $\cal A$. Since $\log\rho \in {\cal A}$, there
exist real numbers $\left\{\theta^1, \ldots, \theta^n, \Psi \right\}$
such that \[\log\rho = \theta^1X_1 + \cdots + \theta^nX_n -
\Psi{\bf 1},\] that is,
\begin{equation}
\rho=\exp\left(\theta^1X_1 + \cdots + \theta^nX_n -
\Psi{\bf 1}\right).
\end{equation}
The normalisation condition $\mbox{Tr}\rho=1$
means, however, that only $n$ among these numbers are independent,
so we choose $\Psi\equiv\Psi(\theta)$ to be the one determined by
the others. Then $\theta = (\theta^1, \ldots, \theta^n)$ form
a $1$-affine coordinate system, as can be seen from the
following calculation
\begin{eqnarray*}
\left(\nabla^{(1)}_{\partial_i}\frac{\partial}{\partial\theta^j}\right)^{(1)}
& = & \frac{\partial^2\log\rho}{\partial\theta^i\partial\theta^j}-
\mbox{Tr}\left(\rho\frac{\partial^2\log\rho}{\partial\theta^i\partial\theta^j}\right)\\
& = & -\frac{\partial^2\Psi}{\partial\theta^i\partial\theta^j}(\theta)
+\mbox{Tr}\left(\rho\frac{\partial^2\Psi}{\partial\theta^i\partial\theta^j}(\theta)\right)\\
& = & 0.
\end{eqnarray*}
Now let ${\cal A}_0$ be the subspace of traceless operators in $\cal A$.
Consider the $-1$-embedding
\begin{eqnarray*}
\ell_{-1} & : & {\cal M} \rightarrow {\cal A} \\
       &   & \rho \mapsto \rho,
\end{eqnarray*}
and define, at each $\rho \in {\cal M}$, the $-1$-representation
of tangent vectors as
\begin{eqnarray*}
(\ell_{-1})_{*(\rho)} & : & T_{\rho}{\cal M} \rightarrow {\cal A}_0 \\
             &   & v \mapsto (\ell_{-1} \circ \gamma)^\prime(0),
\end{eqnarray*}
where $\gamma :(-\varepsilon,\varepsilon)\rightarrow{\cal M}$ is again a
curve in the equivalence class of the tangent vector $v$.
In coordinates, the $-1$-representation of the basis
$\left\{\partial_1,\ldots ,\partial_n\right\}$
of $T_{\rho}{\cal M}$ is $\left\{ \frac{\partial\rho}{\partial\theta^1},
\ldots ,\frac{\partial\rho}{\partial\theta^n}\right\}$. As before, the
$-1$-representation of a vector field $X$ on $\cal M$ is an
${\cal A}_0$-valued function denoted by $(X)^{(-1)}$ or $X^-$.

We obtain the mixture or $-1$-connection by defining the parallel transport
\begin{eqnarray*}
\tau^{(-1)}_{\rho_0,\rho_1} & : & T_{\rho_0}{\cal M} \rightarrow T_{\rho_1}{\cal M}
\\                   &   & v \mapsto
(\ell_{-1})_{*(\rho_1)}^{-1}\left((\ell_{-1})_{*(\rho_0)}v\right),
\end{eqnarray*}
and we find that its covariant derivative in the direction $\partial_i$ is
\begin{equation}
\left(\nabla^{(-1)}_{\partial_i}\frac{\partial}{\partial\theta^j}\right)^{(-1)}=
\frac{\partial^2\rho}{\partial\theta^i\partial\theta^j}.
\end{equation}
If we equip $\cal A$ with the trace norm, then the
$-1$-embedding maps $\cal M$ into the unit sphere $\cal S$ of $\cal A$,
and the $-1$-connection given here is nothing but the projection
onto $\cal S$ of the {\em natural} flat connection in this space. It
turns out that the unit sphere with respect to the trace norm is flat
in  $\cal A$, hence the $-1$-connection is flat on $\cal M$.
Again, it is a convenience of the finite dimensional case that we
can prove flatness of the $-1$-connection by direct construction of a
$-1$-affine coordinate system. Suppose $\left\{X_1, \ldots, X_{n+1}\right\}$
is a normalised basis for $\cal A$, then there exist real numbers $(\xi_1, \ldots, \xi_{n+1})$
such that
\begin{equation}
\rho = \xi_1X_1 + \cdots + \xi_{n+1}X_{n+1}.
\end{equation}
Since $\mbox{Tr}\rho =1$, we can take $\xi_{n+1}= 1 - (\xi_1 + \cdots + \xi_n)$
as a function of the first $n$ independent parameters $\xi_1, \ldots, \xi_{n}$. Then
$\xi = (\xi_1, \ldots, \xi_{n})$ is a $-1$-affine coordinate system for $\cal M$, because
\begin{eqnarray*}
\left(\nabla^{(-1)}_{\frac{\partial}{\partial\xi_i}}\frac{\partial}{\partial\xi_j}\right)^{(-1)}
& = & \frac{\partial^2\rho}{\partial\xi_i\partial\xi_j}\\
& = & \frac{\partial}{\partial\xi_i}(X_j - X_{n+1})\\
& = & 0.
\end{eqnarray*}

\section{Duality and the {\em BKM} Metric}

Two connections $\nabla$ and $\nabla^*$ on a Riemannian manifold $({\cal M},g)$
are dual with respect to $g$ if and only if
\begin{equation}
X g(Y,Z)= g \left(\nabla_X Y,Z \right)+g \left(Y,\nabla^*_X Z\right),
\end{equation}
for any vector fields $X,Y,Z$ on $\cal M$ \cite{Amari1,Murray}. Equivalently,
if $\tau_{\gamma(t)}$ and $\tau^*_{\gamma(t)}$ are the respective parallel
transports along a curve $\{\gamma(t)\}_{0\leq t\leq 1}$ on $\cal M$, with
$\gamma(0)=\rho$, then $\nabla$ and $\nabla^*$ are dual with respect to $g$
if and only if for all $t\in[0,1]$,
\begin{equation}
g_\rho(Y,Z)=g_{\gamma(t)}\left(\tau_{\gamma(t)}Y,\tau^*_{\gamma(t)}Z\right).
\end{equation}

Given any connection $\nabla$ on $({\cal M},g)$, we can always
find a unique connection $\nabla^*$ such that $\nabla$ and
$\nabla^*$ are dual with respect to $g$. On the other hand, given
two connections $\nabla$ and $\nabla^*$, we can ask what are the
possible Riemannian metrics $g$ with respect to which they are
dual. In particular, we want to explore this question for the case
of the exponential and mixture connections on a manifold of
density matrices.

A different concept of duality also used by Amari \cite{Amari1}
is that of dual coordinate systems, regardless of any connection.
Two coordinate systems $\theta = (\theta^i)$ and $\eta = (\eta_i)$ on a Riemannian manifold
$({\cal M},g)$ are dual with respect to $g$ if and only if their
natural bases for $T_\rho{\cal M}$ are {\em biorthogonal} at every point
$\rho\in {\cal M}$, that is,
\[g\left(\frac{\partial}{\partial\theta^i},\frac{\partial}{\partial\eta_j}\right)=\delta^i_j.\]
Equivalently, $\theta = (\theta^i)$ and $\eta = (\eta_i)$ are dual
with respect to $g$ if and only if
\[g_{ij}=\frac{\partial\eta_i}{\partial\theta^j} \quad \mbox{and}
\quad g^{ij}=\frac{\partial\theta_i}{\partial\eta^j}\]
at every point $\rho\in {\cal M}$, where, as usual,
$g^{ij}=(g_{ij})^{-1}$.

The following theorem gives a characterisation of dual coordinate
systems in terms of potential functions, thus bringing
convexity theory and the related duality with respect to Legendre transforms
into the discussion.

\begin{theorem}[Amari, 1985]
When a Riemannian manifold $({\cal M},g)$ has a pair of dual coordinate
systems $(\theta,\eta)$, there exist potential functions $\Psi(\theta)$
and $\Phi(\eta)$ such that
\[g_{ij}(\theta)=\frac{\partial^2\Psi(\theta)}{\partial\theta^i\partial\theta^j}
\quad \mbox{and} \quad
g^{ij}=\frac{\partial^2\Phi(\eta)}{\partial\eta_i\partial\eta_j}.\]
Conversely, when either potential function $\Psi$ or $\Phi$ exists
from which the metric is derived by differentiating it twice, there
exist a pair of dual coordinate systems. The dual
coordinate systems and the potential functions are related by the following Legendre
transforms
\[\theta^i=\frac{\partial\Phi(\eta)}{\partial\eta_i},
\quad \eta_i=\frac{\partial\Psi(\theta)}{\partial\theta^i}\]
and
\[\Psi(\theta)+\Phi(\eta)-\theta^i\eta_i = 0\]
\label{potentials}
\end{theorem}

In contrast to the case of dual connections, dual coordinate
systems do not necessarily exist on every Riemannian manifold \cite{Amari1}.
When the additional property of flatness is required,
the following theorem provides a link between the two concepts of duality.
In the sense used in this paper, a connection $\nabla$ on manifold $\cal M$ is
said to be flat if $\cal M$ admits a global $\nabla$-affine
coordinate system. This is equivalent to its curvature and
torsion both being zero.

\begin{theorem}[Amari, 1985]
Suppose that $\nabla$ and $\nabla^*$ are two flat connections on a manifold $\cal M$.
If they are dual with respect to a Riemannian metric $g$ on
$\cal M$, then there exists a pair $(\theta,\eta)$
of dual coordinate systems such that $\theta$ is $\nabla$-affine
and $\eta$ is a $\nabla^*$-affine.
\label{dualities}
\end{theorem}

We now return to our manifold $\cal M$ of density matrices and
consider the problem of finding a unique Riemannian metric for it.
Using either the $1$ or the $-1$ representation of the
tangent bundle $T{\cal M}$, we define a Riemannian metric on $\cal M$
by a smooth assignment of an inner product $\langle\cdot,\cdot\rangle_
{\rho}$ in
${\cal A}\subset B({\cal H}^N)$ for each point $\rho \in {\cal M}$.
The {\em BKM} (Bogoliubov-Kubo-Mori) metric is the Riemannian metric on
$\cal M$ obtained from the {\em BKM} inner product \cite{Petz2}.
If $A^{(1)},B^{(1)}$ and
$A^{(-1)},B^{(-1)}$ are, respectively, the $1$ and $-1$ representations of
$A,B \in T_{\rho}{\cal M}$, then the {\em BKM} metric has the following
equivalent expressions:
\begin{eqnarray}
g^{\scriptscriptstyle B}_{\rho}(A,B) & = & \mbox{Tr}\left(A^{(-1)}
B^{(1)}\right)\nonumber\\
& = & \int_{0}^{1}\mbox{Tr}\left(\rho^{\lambda}A^{(1)}\rho^{1-\lambda}
B^{(1)}\right)d\lambda\nonumber\\
& = & \int_{0}^{\infty}\mbox{Tr}\left(\frac{1}{t+\rho}A^{(-1)}\frac{1}
{t+\rho}B^{(-1)}\right)dt\label{bkm}
\end{eqnarray}

It is well known that the exponential and mixture connections are
dual with respect to the {\em BKM} metric \cite{Nagaoka,Hasegawa2}.
A natural question is whether it is the unique metric with this
property. The next theorem tells us how much uniqueness can be
achieved from duality alone.

\begin{theorem}
If the connections $\nabla^{(1)}$ and $\nabla^{(-1)}$ are dual
with respect to a Riemannian metric $g$ on $\cal M$, then there
exist a constant (independent of $\rho$) $n\times n$ matrix $M$,
such that $(g_{\rho})_{ij}={\displaystyle \sum_{k=1}^{n}}
M_i^k(g^{\scriptscriptstyle B}_{\rho})_{kj}$, in some $1$-affine
coordinate system. \label{unique1}
\end{theorem}
{\em Proof:} Since the two connections are flat, by
theorem~\ref{dualities}, there exist dual coordinate systems $(\theta,\eta)$
such that $\theta$ is $\nabla^{(1)}$-affine and $\eta$ is
$\nabla^{(-1)}$-affine. Thus, applying theorem~\ref{potentials},
there exist a potential function $\Psi(\theta)$ such that
\[g_{ij}(\theta)=\frac{\partial^2\Psi(\theta)}{\partial\theta^i\partial\theta^j}\]
and
\[\eta_i=\frac{\partial\Psi(\theta)}{\partial\theta^i}.\]
On the other hand, since $\theta$ is $\nabla^{(1)}$-affine, there
exist linearly independent operators\\
$\left\{{\bf{1}}, X_1, \ldots, X_n\right\}$ such that
\begin{equation}
\rho=\exp\left(\theta^1X_1 + \cdots + \theta^nX_n -
\tilde{\Psi}(\theta){\bf 1}\right)
\label{rho},
\end{equation}
where
\begin{equation}
\tilde{\Psi}(\theta)=\log\mbox{Tr}\left[\exp\left(\theta^1X_1 + \cdots +
\theta^nX_n\right)\right].
\label{free-energy}
\end{equation}
Without loss of generality, we can assume that the operators $X_1, \ldots, X_n$
are traceless, since if we add multiples of the identity to any $X_j$
in~(\ref{rho}), we can still have the same parameters $\theta$ as coordinates
for the same point $\rho$ just by modifying the function $\tilde{\Psi}$. But any such set
of operators define a $\nabla^{(-1)}$-affine coordinate system through the formula
\[\tilde{\eta}_i=\mbox{Tr}(\rho X_i),\] because the latter are affinely related to the $\xi$ coordinates defined in section 2 (with $X_{n+1}={\bf 1}/n$).
Differentiating $\tilde{\Psi}$ with respect to $\theta^i$ we obtain
\[\frac{\partial\tilde{\Psi}(\theta)}{\partial\theta^i}=\mbox{Tr}(\rho X_i)=
\tilde{\eta}_i.\]
Thus $\tilde{\eta}_i=\frac{\partial\tilde{\Psi}(\theta)}{\partial\theta^i}$ and
$\eta_i=\frac{\partial\Psi(\theta)}{\partial\theta^i}$
are two $\nabla^{(-1)}$-affine coordinate systems, so they must be
related by an affine transformation \cite{Murray}. So there
exist an $n\times n$ matrix $M$ and numbers $(a_1, \ldots, a_n)$
such that
\[\eta_i = \sum_{k=1}^{n}M_i^k\tilde{\eta}_k + a_i,\]
that is,
\[\frac{\partial\Psi(\theta)}{\partial\theta^i} =
\sum_{k=1}^{n}M_i^k\frac{\partial\Psi(\theta)}{\partial\theta^k} +
a_i,\]
and differentiating this equation  with respect to $\theta^j$ gives
\begin{equation}
g_{ij}(\theta)=\frac{\partial^2\Psi(\theta)}{\partial\theta^i\partial\theta^j}
=
\sum_{k=1}^{n}M_i^k
\frac{\partial^2\tilde{\Psi}(\theta)}{\partial\theta^j\partial\theta^k}.
\label{psis}
\end{equation}
But we can calculate the second derivative of $\tilde{\Psi}$ directly
from~(\ref{free-energy}), obtaining
\begin{eqnarray*}
\frac{\partial^2\tilde{\Psi}(\theta)}{\partial\theta^j\partial\theta^k} &
= & \int_{0}^{1}\mbox{Tr}\left(\rho^{\lambda}\frac{\partial\log\rho}{\partial\theta^j}
\rho^{1-\lambda}\frac{\partial\log\rho}{\partial\theta^k}\right)d\lambda\\
& = & g^{\scriptscriptstyle B}_{\rho}
\left(\frac{\partial}{\partial\theta^j},\frac{\partial}{\partial\theta^k}\right)\\
& = & g^{\scriptscriptstyle B}_{jk}(\theta).
\end{eqnarray*}
Inserting this in~(\ref{psis}) completes the
proof.\hspace{\fill}$\Box$

\section{The Condition of Monotonicity}

We have seen in the previous section that requiring duality
between the exponential and mixture connections reduces the set of
possible Riemannian metrics on $\cal M$ to matrix multiples of
the {\em BKM} metric. We now investigate the effect of imposing a
monotonicity property on this set.

If we use the $-1$-representation to define a Riemannian metric $g$ on $\cal M$
by means of the inner product
$\langle\cdot,\cdot\rangle_{\rho}$ in ${\cal A}\subset B({\cal H}^N)$,
then we say that $g$ is {\em monotone} if and only if
\begin{equation}
\left\langle S(A^{(-1)}),S(A^{(-1)})\right\rangle_{S(\rho)} \leq \left\langle
A^{(-1)},A^{(-1)}\right\rangle_{\rho},
\end{equation}
for every $\rho \in {\cal M}$, $A \in T_{\rho}{\cal M}$, and every
completely positive, trace preserving map $S:{\cal A} \rightarrow {\cal
A}$.

In a series of papers \cite{Petz1,Petz3,PetzSudar}, Petz has given
a complete characterisation of monotone metrics on full matrix
spaces in terms of operator monotone functions. The Chentsov
condition, however, is defined for metrics on the space ${\cal M}$
of faithful {\em states}, and must first be extended to ${\cal A}$
before we can use Petz's results. Let $\widehat{\cal M}$ be the
manifold of faithful weights (the positive definite matrices). Let
$g$ be a metric on $T{\cal M}$. We can extend $g$ to ${\cal
A}\simeq T\widehat{\cal M}$ as follows. At $\rho\in{\cal M}$ and
$\hat{A},\hat{B} \in T\widehat{\cal M}_\rho$, put
$\hat{A}^{(-1)}=A_0\rho+A^-$, where $A_0=\mbox{Tr}\,\hat{A}^{(-1)}
\in{\bf R}$ and $\mbox{Tr}\,A^-=0$, and similarly for $B$. Then
put
\begin{equation}
\hat{g}_\rho(\hat{A},\hat{B}):=A_0B_0+g_\rho(A^-,B^-). \label{ext}
\end{equation}
For $g^{\scriptscriptstyle B}$ this extension coincides with that given by
eq.~(\ref{bkm}). More generally, if $g$ is monotone on $T{\cal M}$, then
$\hat{g}$ is monotone on $T\widehat{\cal M}$. For, let $S$ be a
trace-preserving completely positive map on $T\widehat{\cal M}$, and $\rho\in{\cal M}$.
Then $S$ maps $T{\cal M}$ to itself, and
\[
\hat{g}_{_{S\rho}}(S\hat{A},S\hat{A})=A_0^2+g_{_{S\rho}}(SA^-,SA^-)\leq
A_0^2+ g_\rho(A^-,A^-)=\hat{g}_\rho(A,A).\]

For any metric $\hat{g}$ on $T\widehat{\cal M}$, and putting
$\hat{A}^{(-1)}= A_0\rho+A^-$, we define the positive (super)
operator $K_\rho$ on $\cal A$ by
\begin{equation}
\hat{g}_\rho(\hat{A},\hat{B})=\left\langle
\hat{A}^{(-1)},K_\rho\left(\hat{B}^{(-1)}\right)
\right\rangle_{\scriptscriptstyle H \!
S}=\mbox{Tr}\left(\hat{A}^{(-1)}K_\rho
\left(\hat{B}^{(-1)}\right)\right). \label{metric}
\end{equation}
Note that our $K$ is denoted $K^{-1}$ by Petz. He also defines the
(super) operators, $L_\rho X:=\rho X$ and $R_\rho X:=X\rho$, for
$X \in {\cal A}$, which are also positive. Then he proved
\cite{Petz3}

\begin{theorem}[Petz]
A Riemannian metric $g$ on $\cal A$ is monotone if and only if
\[K_{\rho}=\left(R_{\rho}^{1/2}f(L_{\rho}R_{\rho}^{-1})R_{\rho}^{1/2}
\right)^{-1},\]
where $K_{\rho}$ is defined in~(\ref{metric}) and $f:R^{+}\rightarrow R^{+}$
is an operator monotone function satisfying $f(t)=t f(t^{-1}).$
\label{function}
\end{theorem}

In particular, the {\em BKM} metric is monotone and its corresponding operator
monotone function is
\[f^{\scriptscriptstyle B}(t)=\frac{t-1}{\log t}.\]

Combining this characterisation with our theorem~(\ref{unique1}), we
obtain the following improved uniqueness result.

\begin{theorem}
If the connections $\nabla^{(1)}$ and $\nabla^{(-1)}$ are dual with respect
to a monotone Riemannian metric $g$ on $\cal M$, then $g$ is a
{\em scalar} multiple of the BKM metric.
\end{theorem}
{\em Proof:} We first extend the $(N^2-1)\times(N^2-1)$ matrix $M$ of
theorem~(\ref{unique1}) by one row and column, to a matrix
$\hat{M}$, where $\hat{M}_{ij}=M_{ij},\;(1\leq i,j\leq N^2-1)$,
and
\[
\hat{M}_{i,N^2}=\hat{M}_{N^2,i}=\delta_{i,N^2},
(i=1,\ldots,N^2).\] Now, in the coordinates
$\hat{A}^{(-1)}=A_0\rho+A^-$ of eq.~(\ref{ext}), the extension of
$g$ to $\hat{g}$ in the orthogonal direction $\rho$ is the same
for all $g$. In particular,
\[\hat{g}(A_0\rho,B^-)=\hat{g}^{\scriptscriptstyle B}(A_0\rho,B^-)=0;\hspace{.5in}
\hat{g}(A_0\rho,B_0\rho)=\hat{g}^{\scriptscriptstyle B}(A_0\rho,B_0\rho)=A_0B_0.\]
It follows that the equation proved in theorem~(\ref{unique1}), when expressed in the $-1$-affine coordinates $\eta$ (by taking inverses and using
 theorem~\ref{potentials}), gives the following relation between the kernels of $\hat{g}$ and $\hat{g}^{\scriptscriptstyle B}$, say  $K^g$ and
$K^{\scriptscriptstyle B}$ respectively,
\begin{equation}
K^g_{\rho}=\hat{M}^{\scriptstyle t}K^{\scriptscriptstyle B}_{\rho}.
\end{equation}
Therefore, if $f^g$ and $f^{\scriptscriptstyle B}$ are the operator
monotone functions
corresponding respectively to $g$ and $g^{\scriptscriptstyle B}$, from
theorem~\ref{function}, we have
\begin{eqnarray*}
\left(R_{\rho}^{1/2}f^g(L_{\rho}R_{\rho}^{-1})R_{\rho}^{1/2}\right)^{-1}
& = & \hat{M}^{\scriptstyle t}\left(R_{\rho}^{1/2}f^{\scriptscriptstyle B\! K\! M}
(L_{\rho}R_{\rho}^{-1})R_{\rho}^{1/2}\right)^{-1}\\
\left(R_{\rho}^{1/2}f^g(L_{\rho}R_{\rho}^{-1})R_{\rho}^{1/2}\right)\hat{M}^{\scriptstyle t}
& = & \left(R_{\rho}^{1/2}f^{\scriptscriptstyle B\! K\! M}
(L_{\rho}R_{\rho}^{-1})R_{\rho}^{1/2}\right)\\ \hat{M}^{\scriptstyle t} & = &
f^g(L_{\rho}R_{\rho}^{-1})^{-1}f^{\scriptscriptstyle B\! K\! M}
(L_{\rho}R_{\rho}^{-1}),
\end{eqnarray*}
as everything commutes.
Thus, the operator $\hat{M}$ is given as a function of the operator
$L_{\rho}R_\rho^{-1}$, but it is itself independent of the point $\rho$,
so we conclude that it must be a scalar multiple of the identity
operator.\hspace{\fill}$\Box$

\section{Discussion}

We have proved that, in a finite dimensional quantum system, the
{\em BKM} metric is, up to a scalar factor, the unique monotone
Riemannian metric with respect to
which the exponential and mixture connections are dual. It should
be emphasised that we have considered the manifold of {\em all}
invertible density operators. The uniqueness result might not be
true for a submanifold of $\cal M$, such as occurs in parametric estimation
theory.

We should compare our uniqueness theorem with the results of
Hasegawa \cite{Hasegawa3} and Petz  and Hasegawa
\cite{HasegawaPetz,PetzHasegawa}, who obtained a family of metrics
$\{g^{\alpha}\}$, each corresponding to a divergence based on the
corresponding Wigner-Yanase information, as generalised by Dyson.
They show that these metrics are monotone, and that the
$\alpha$-divergence `admits dual connections', in a sense they
specify. This does not contradict our result; the $\pm
1$-connections are not dual relative to $g^{\alpha}$ unless
$\alpha=\pm 1$, which is the {\em BKM} case.

Having fixed the metric, the next step in the development of the theory is to
define the $\alpha$-connections, for $\alpha \in (-1,1)$. We  look for a
definition that
makes $\nabla^{(\alpha)}$ and $\nabla^{(-\alpha)}$ dual and such that
the extended manifold
$\widehat{\cal M}$ is $\alpha$-flat. Only then we can try to find the
quantum analogue of the $\alpha$-divergence. Duality is easily
achieved if we define the $\alpha$-connections as the convex
mixture $\nabla^{(\alpha)}=\frac{1+\alpha}{2}\nabla^{(1)} +
\frac{1-\alpha}{2}\nabla^{(-1)}$.
Flatness, on the other hand, is more apparent if we use the
$\alpha$-embedding. We have not yet been able to prove that the
two definitions are equivalent. All these questions are deferred to
a subsequent paper.


\begin{thebibliography}{99}

\bibitem{Amari1} S.-I. Amari, {\bf Differential Geometric Methods in
Statistics}, {\em Lecture Notes in Statistics}, {\bf 28}, Springer-Verlag,
New York, 1985.

\bibitem{Amari2} S.-I. Amari, {\em Information Geometry}, Contemp. Math., {\bf 203},
1985, pp. 81-95.

\bibitem{Bogoliubov} N. N. Bogoliubov, Phys. Abh. Sov. Union,
{\bf 1}, 229, 1962.

\bibitem{Chentsov} N. N. \v{C}encov, {\bf Statistical Decision Rules
and Optimal Inferences}, {\em Translations of Mathematical
Monographs}, American Mathematical Society, Providence, 1982.

\bibitem{GibiliscoIsola} P. Gibilisco and T. Isola, {\em
Connections on Statistical Manifolds of Density Operators by Geometry of
Noncommutative $L^p$-spaces}, Inf. Dim. Analysis, Quant. Prob. and Rel. Top.,
{\bf 2}, 169-178, 1999.

\bibitem{GrasselliStreater} M. R. Grasselli and R. F. Streater, {\em The Quantum
Information Manifold for $\varepsilon$-bounded Forms}, to appear in Rep. Math.
Phys., math-hp/9910031.

\bibitem{Hasegawa1} H. Hasegawa, {\em Noncommutative Extension of
the Information Geometry}, in {\bf Quantum Communication and
Measurement}, eds. V. P. Balavkin, O. Hirota and R. L. Hudson,
Plenum Press, 1995; pp. 327-337.

\bibitem{Hasegawa2} H. Hasegawa, {\em Exponential and mixture families
in quantum statistics: dual structures and unbiased parameter estimation.},
Rep. Math. Phys., {\bf 39}, 49-68, 1997.

\bibitem{Hasegawa3} H. Hasegawa, {\em $\alpha$-divergence of the
non-commutative information geometry}, Rep. Math. Phys., {\bf 33}, 87-93,
1993.

\bibitem{HasegawaPetz} H. Hasegawa and D. Petz, {\em Non-commutative
extension of information geometry II}, in {\bf Quantum Communication
and Measurement}, eds. V. P. Belavkin, O. Hirota and R. L. Hudson,
Plenum Press, pp 109-118, 1997.

\bibitem{Kubo} R. Kubo, {\em Statistical Mechanical Theory of
Irreversible Processes I}, J. Phys. Soc. Japan, {\bf 12}, 570, 1957, also
{\em The Fluctuation-Dissipation Theorem}, Rep. Prog. Phys, {\bf 29}, 255,
1966.

\bibitem{Mori} H. Mori, {\em Transport, Collective Motion and
Brownian Motion}, Prog. Theor. Phys., {\bf 33}, 423, 1965.

\bibitem{MorozovaChentsov} E. A. Morozova and N. N. Chentsov, {\em
Markov Invariant Geometry on State Manifolds} (Russian), Ito.
Nauki i Tek., {\bf 36}, 69-102, 1990.

\bibitem{Murray} M. K. Murray and J. W. Rice, {\bf Differential
Geometry and Statistics}, {\em Monographs on Statistics and Applied
Probability}, {\bf 48}, Chapman \& Hall, 1993.

\bibitem{Nagaoka} H. Nagaoka, {\em Differential Geometric Aspects
of Quantum State Estimation and Relative Entropy}, in {\bf Quantum Communication and
Measurement}, eds. V. P. Balavkin, O. Hirota and R. L. Hudson,
Plenum Press, 1995.

\bibitem{Petz1} D. Petz, {\em Quasi-entropies for Finite Quantum Systems},
Rep. Math. Phys., {\bf 23}, 57-65, 1986.

\bibitem{Petz2} D. Petz, {\em Geometry of Canonical Correlation on
the State Space of a Quantum System}, J. Math. Phys., {\bf 35},
780-795, 1994.

\bibitem{Petz3} D. Petz, {\em Monotone Metrics on Matrix Spaces}, Lin. Alg. Appl.,
{\bf 244}, 81-96, 1996.

\bibitem{PetzHasegawa} D. Petz and H. Hasegawa, {\em On the Riemannian
metric of $\alpha$-entropies of density matrices}, Lett. Math. Phys.,
{\bf 38}, 221-241, 1996.

\bibitem{PetzSudar} D. Petz and C.Sudar, {\em Geometries of Quantum
States}, J. Math. Phys., {\bf 37}, 2662-2673, 1996.


\bibitem{Streater} R. F. Streater, {\em The Information Manifold
for Relatively Bounded Potentials}, Proc. Stecklov Institute
Maths. {\bf 228}, 205-223, 2000, math-ph/9910035.




\end{thebibliography}
\end{document}